\documentclass{article}
\usepackage[utf8]{inputenc}

\usepackage{amsmath, graphicx}

\usepackage{xcolor}
\usepackage[numbers]{natbib}
\title{Diversity in Valuing Social Contact and Risk Tolerance Lead to the Emergence of Homophily in Populations Facing Infectious Threats}
\author{Matthew J. Young, Matthew J. Silk, Alex J. Pritchard, Nina H. Fefferman}
\begin{document}

\maketitle

\section{Abstract}
How self-organization leads to the emergence of structure in social populations remains a fascinating and open question in the study of complex systems. One frequently observed structure that emerges again and again across systems is that of self-similar community, \textit{i.e.}, homophily. We use a game theoretic perspective to explore a case in which individuals choose affiliation partnerships based on only two factors: the value they place on having social contacts, and their risk tolerance for exposure to threat derived from social contact (e.g., infectious disease, threatening ideas, etc.). We show how diversity along just these two influences are sufficient to cause the emergence of self-organizing homophily in the population. We further consider a case in which extrinsic social factors influence the desire to maintain particular social ties, and show the robustness of emergent homophilic patterns to these additional influences. These results demonstrate how observable population-level homophily may arise out of individual behaviors that balance the value of social contacts against the potential risks associated with those contacts. We present and discuss these results in the context of outbreaks of infectious disease in human populations. Complementing the standard narrative about how social division alters epidemiological risk, we here show how epidemiological risk may deepen social divisions in human populations.

\section{Introduction}
Many studies have considered the impact of individual homophilic behaviors on emergent social structures \cite{fu2012evolution,currarini2016simple, treur2019mathematical,holme2006coevolution,pearson2006homophily,mcpherson2001homophily}. Communities in which individuals are more likely to interact with others like them have been discussed widely as both positives (e.g. modularity potentially inhibiting disease spread \cite{sah2017unraveling,salathe2010dynamics,evans2021group}) and negatives (e.g. echo chambers into which important information may be less able to penetrate \cite{cinelli2021echo}). Few studies, however, have considered whether observable homophily in population organizational structure can itself have emerged from more fundamental self-organizing individual-level behaviors. For members of social species, there is a likely benefit from social contacts themselves, but also that the benefit may be dampened by costs associated with risk that arises directly from that same contact. One clear example of such systems is infectious disease spreading through social populations.

During a disease outbreak, infection risk can be mitigated through preventative measures \cite{leppin2009risk}, such as social distancing, that have been presented as part of a behavioral immune system \cite{schaller2011behavioural,schaller2006parasites}. Individual adherence to preventative behaviors has been associated with experimental measures of risk aversion \cite{collier2020risk}. Thus, all else equal, in response to an epidemic, more risk averse individuals can be expected to adhere more to social distancing. By definition, the act of social distancing impedes interactions; a phenomenon that is antithetical with human beings’ need-to-belong \cite{baumeister1995need}. Loneliness can have a diversity of health and mental costs \cite{hawkley2010loneliness}; pain associated with social circumstances shares neural pathways with physical pain \cite{eisenberger2004rejection}. This trade-off between the need-to-belong and the behavioral immune system has not gone unrecognized, and has been the subject of several laboratory-based studies \cite{sawada2018activation, sacco2014balancing}. Outside of the laboratory, however, individuals live in a heterogeneous society with distinct predilections towards prioritizing social interactions against their own risk aversion.

Explicit consideration of this social heterogeneity can be explored by incorporating individually distinct strategies of decision-making. Both social isolation and infection have the capacity to cause very real physical and mental maladies; as such, there is unlikely to be a unique, globally optimal solution that accommodates needs across a diverse population. Contingent on risk categories (such as age, immunocompetence, nutrition, or underlying health factors; \cite{biswas2020effects, maggini2018immune}), individuals can face uncertainty as to the actual risks of contracting a disease and its severity, given those contingencies. This is significant, as risk-averse individuals have been suggested to be intolerant, or aversive, of uncertainty \cite{caligiuri2012dynamic, lauriola2007common, petrocchi2021interplay}. Importantly, risk assessment and aversion have a deep evolutionary history \cite{blanchard2011riskassess, smith2008fitness}, with evidence for them to be heritable traits \cite{vanoers2003heritab, cesarini2009genetic}. In humans, individual differences, such as the extraversion or sociability dimensions of personality, have been shown to predict scores on scales of risk-taking behavior, inclusive of health behavior \cite{zuckerman2001risk-taking, nicholson2005persandrisk}. An infectious disease outbreak simultaneously introduces large-scale (often unequally perceived) risk factors and, thus, has the capacity to alter the social dynamics of a community \cite{van2020using,pagetinfluenza}. 

Risk aversion during a disease outbreak has been posited to be an ‘impure public good’ \cite{collier2020risk,baumeister1995need,fenichel2013economic,bell2009macroeconomics}: individuals’ actions contribute to public health, but individuals may not have a vested interest in public health (either actual or self-perceived) and are, instead, acting out of self-preservation \cite{fenichel2013economic, bell2009macroeconomics}. Such a perspective emphasizes the importance of understanding how individuals’ personal decision-making processes, in the midst of an  outbreak, scale to have emergent social consequences. These emergent social consequences could have important implications for the progression of the outbreak while simultaneously altering people's experience of their social world and the benefits that they obtain from socializing. One such emergent consequence might be that individuals preferentially associate with alike individuals, which might be evident via social assortativity, i.e., homophily \cite{mcpherson2001homophily}. Risk aversion has previously been recognized as a key component of the structural emergence of homophily \cite{currarini2016identity, kovavrik2014risk}. For example, in an economic context, models of the generation of network structure  have prioritized risk aversion \cite{kovavrik2014risk}.  
 
 Here we present a game theoretical approach to understanding how prioritization of individual strategies may shape the social associations of a population without a preexisting social structure (i.e., lacking a pre-imposed social network or hierarchy). We implement a socializing game, in which individuals have distinct strategies for prioritizing risk or socialization and demonstrate how these simple dynamics may lead to well-defined emergent structures that have implications for both disease risk and societal function.

\section{Methods}

We define a socializing game during an epidemic with susceptible-infected-susceptible (SIS) dynamics.  We assume players are short-sighted and subrational; rather than being perfectly rational agents who can predict the behavior of other players and compute the optimal response, players repeatedly play games and incrementally update their behavior in the direction of the best response to their current environment.

\subsection*{Game Formalism}

Let $M >> 0$ be an arbitrary large number. At each time $t$, for each player $y$, player $x$ decides on a social action strategy $\vec{a}_x(t) = \{a_{x,y}(t) \in [0,M]: y \in G\}, $ corresponding to an amount of social contact they want to have with each other player. \footnote{This assumes that all players in the population are socially connected to each other.  If we relax this assumption by arranging players spatially and setting $a_{x,y} = 0$ for players more than a certain distance apart then, assuming players from each group are equally distributed spatially and so is the initial infection, this will yield identical results to our model.  If we don't make those assumptions, or relax this restriction in other ways such as arranging players on a social network graph, then this is likely to create new dynamics, but is beyond the scope of this work.}  This social contact requires mutual agreement, so the amount of socializing that actually occurs between players $x$ and $y$ is
$$
a'_{x,y}(t) := min(a_{x,y}(t),a_{y,x}(t))
$$

The utility of player $x$ at time $t$ is defined by

$$
u_x(t) = \sigma_x f(a_x) - \rho_x p(\vec{a'}_x)
$$

Where $f$ is a nondecreasing concave function (such as $ln$) which yields greater utility for higher inputs at diminishing rates,  $a_x = \int a'_{x,y} dy$, and $p(\vec{a'}_x)$ is the expected probability of becoming infected as a result of socializing per unit of time.

\subsection*{Epidemiological Calculations within the Game Formalism}

Since the force of infection can be expected to scale proportionally to the amount of social contact among the players \cite{arregui2018projecting}, then, assuming player $x$ is uninfected,

$$
p(\vec{a'}_x) = \int \beta  a'_{x,y} q_y dy
$$

where $\beta$ be the inherent transmissibility of the disease, and $q_y$ is $1$ if player $y$ is infected, and $0$ otherwise.

Let $\gamma$ be the recovery rate of an infected individual from the disease, and assume that recovered individuals return to the susceptible population with no residual protective immunity.  %Let $I_k$ be the proportion of group $k$ that is infected.  
Then this defines a Susceptible-Infectious-Susceptible (SIS) model \cite{anderson1992infectious} with

$$
\frac{dI}{dt} = -I \gamma + \int (1-q_x) p(\vec{a'}_x) dx
$$

where we explicitly model the contribution to the $I$ class as the expected rate of recruitment determined by the behaviorally-driven force of infection.

\subsection*{Logic of the System}

To simplify simulation and analysis of this system, we consider populations of players with $\sigma_x$ and $\rho_x$ determined randomly from discrete probability distributions.  We define $\sigma_l < \sigma_h$, $\rho_l < \rho_h$, and assume each player $x$ has an equal chance of having $\sigma_x$ equal $\sigma_l$ or $\sigma_h$, and $\rho_x$ equal $\rho_l$ or $\rho_h$.  This effectively partitions the set of all players $G$ into four homogeneous subsets of equal size $G_j$.  Because all players within a group have the same utility function, they will make the same decision regarding $\vec{a}$. When referring to specific groups, we denote them identifiers based on their combined parameters: AS (asocial, $\sigma_j = \sigma_l$), SO (social, $\sigma_j = \sigma_h$), RT (Risk-Taking, $\rho_j = \rho_l$), RA (Risk-Averse, $\rho_j = \rho_h$).  Combining these identifiers leads to the four subgroups: ASRT, ASRA, SORT, SORA.

Additionally, we assume that the utility functions of each player is common knowledge.  We also assume players know the frequency of infection within each subgroup, but do not know the infection status of any particular individual.  So they can, for instance, choose to socialize more with groups that currently have fewer infected members, but cannot choose to socialize exclusively with uninfected individuals within a group. (Note: this is consistent with unreliable proximate cues of infection, especially for diseases with a pre-symptomatic phase; \cite{catching2021examining, cetrulo2020privilege, townsend2020emerging}.)

Due to this information restriction, we can rewrite all variables using group-subscripts.  The amount of socializing among these groups will be

\begin{eqnarray*}
    & & a_{j,k}(t) = |G_k| a_{x,y}(t) \\
    & & a'_{j,k}(t) = |G_k| a'_{x,y}(t)
\end{eqnarray*}

where $x$ is any player from $G_j$, and $y$ is any player from $G_k$.  This allows us to write the SIS dynamics in terms of these subgroups.  Let $I_j(t)$ be the fraction of players in $G_j$ who are infected at time $t$.  Then

%group SIS

$$
\frac{dI_j}{dt} = -I_j \gamma + \sum_k (1-I_j) I_k \beta a'_{j,k}
$$

and likewise the probability of becoming infected for an uninfected member of group $j$ is

$$
p(\vec{a'}_j) = \sum_k I_k \beta a'_{j,k}
$$

If we let the socialization function $f(a) = ln(a)$, and substitute, the utility function for a member of group $j$ is

$$
u_j(t) = \sigma_j ln(\sum_k a'_{j,k}) - \rho_j \sum_k I_k \beta a'_{j,k}
$$

\subsection*{Population Dynamics}

Rather than rationally predicting other player's strategies and computing the optimal response to every circumstance, players gradually adapt their strategy over time in response to feedback from games they play (i.e., subrationality).  In particular, each player starts with a set of strategies $\vec{a}$ corresponding to their pre-existing social preferences.  At each time step, players play one round of the game with each other, and receive payoffs according to the results.  Some time $\Delta t$ passes, and the disease progresses according to the SIS equations and the amount of socialization played in the game.  Each player is then informed of the new frequency of infection $I_k$ for each subgroup of players.  Each player then updates their social preferences in the direction of best response.  In particular, we fix an update size $\Delta a$, then for each group $k$, player $j$ estimates $u_j$ in the counterfactual situations where their socialization with group $k$ had been $a'_{j,k}+n\Delta a$, where $n \in \{-1,0,1\}$, and all other socialization levels remained unchanged.  The player then increments $a_{j,k}$ by $n\Delta a$ according to the $n$ that yields the highest utility.  This is computed simultaneously and independently for each $k$. \footnote{Because $a_{j,k}$ does not directly control $a'_{j,k}$ except through mutual agreement via $a_{k,j}$, this can lead to a situation in which a player repeatedly increases $a_{j,k}$, but this yields no change in the actual dynamics.  A value of $a_{j,k}$ much greater than $a_{k,j}$ behaves no differently from a value of $a_{j,k}$ that's only slightly greater than $a_{k,j}$ except that it takes longer to update if the incentives change causing player $j$ to start decreasing it.  To prevent cases such as this, we fix a small integer $n_c = 4$ and restrict $a_{j,k} \leq a_{k,j} + n_c \Delta a$}

This allows players to incrementally update their strategy in response to the changing environment of their peer's behavior and the state of the epidemic.  By adjusting the ratio $\frac{\Delta a}{\Delta t}$ we can change the speed of social adaptation relative to the speed of the epidemic.

We assume infected individuals have a latent period in which they are contagious but pre-symptomatic, so they continue socializing at the same rate as uninfected individuals.

\subsection*{Extending the Model to Include Differentially Valued Contacts}

Thus far, the model has remained agnostic among social contacts, allowing any interaction to serve equally to fulfill any individual's desire to socialize. To incorporate scenarios in which certain social contacts across groups are of greater personal value than others, we also fix a constant $m$ corresponding to an additional utility gained from each individual's most high-valued relationships and social activities, and an exponential decay term $r < 1$.  We then incorporate a distribution of utilities across interactions by multiplying the value of each marginal social interaction by $1+m r^{x}$ and instead use  
$$
a_j := c \sum_k \int_{0}^{a'_{j,k}} 1 + m r^x dx
$$
as the term characterizing the social contribution to the utility function, where $n$ is a constant to scale the resulting function.

This simplifies to

$$
a_j = c \sum_k \left( a'_{j,k} + \frac{m(1-r^{a'_{j,k}})}{-ln(r)} \right)
$$

Changing the social contribution of the utility function in this way allows for the maintenance of social contacts across groups and the creation of new ones by creating an incentive for players to spread their social interactions among all of the groups.  Of course, the benefits of socializing with members of different groups are then balanced against the differing infection exposure that results.

\subsection*{Metrics and Analysis}

We record emergent patterns by quantifying social contacts within and across preference groups over time and the coupled outcomes in the prevalence of infection in those groups. 

In order to characterize emergent homophily in the context of this composite utility behavior, we fix constants $w$ and $\tau$ with the goal of fixing $\int_{0}^{\tau} 1 + m r^x dx = w$.
For simulations presented here we use $\tau = 1, w = 1.25$. We let $r$ range from 0 to 1 as a free variable, and set $m = \frac{ln(r)(w/r -\tau)}{r^\tau-1}$. In this way, the area under the curve can be normalized by setting $c = r$.  When $r = 1$ this yields a flat line, equivalent to the model with no social differentiation in value among individual contacts.  As $r$ decreases, $m$ increases, and the ratio in benefit skews in favor of diversity among contacts across groups (i.e., allowing maintenance of particular contacts, despite mismatch in axes of socialization and risk tolerance).  As $r$ goes to 0, the function approaches an infinitely steep exponential decay, which incentivizes socializing across descriptive groups.
We then measure homophily using a modified version of Krackhardt's EI homophily index.  \footnote{The standard EI homophily index uses total number of internal and external network connections, while we define them based on total socialization for our groups.  Additionally, we define E using the average socialization with each external group rather than the sum, so that we get a score of 0 when players are indifferent among socializing among groups, rather than when precisely half of socialization is internal.  Finally, we negate our measure relative to the typical EI homophily score so that perfect homophily yields a score of $1$ rather than $-1$.}  In particular, for each group j we define
\begin{eqnarray*}
    & & I_j = a'_{j,j}\\
    & & E_j = \frac{1}{3}\sum_{k \neq j} a'_{j,k}\\
    & & H_j = \frac{I-E}{I+E}
\end{eqnarray*}

Where $H_j$ measures the homophily in the socialization of each group.  We then define the overall homophily of the population, $H$, as the average $H_j$.  When players socializing equally among all groups, we get $H = 0$, as players socialize more in their own groups this increases up to a maximum of $1$, and in the hypothetical scenario where players socialized more externally this decreases down to a minimum of $-1$.

\section{Results and Discussion}

\begin{figure}[h]
\includegraphics[scale = 0.35]{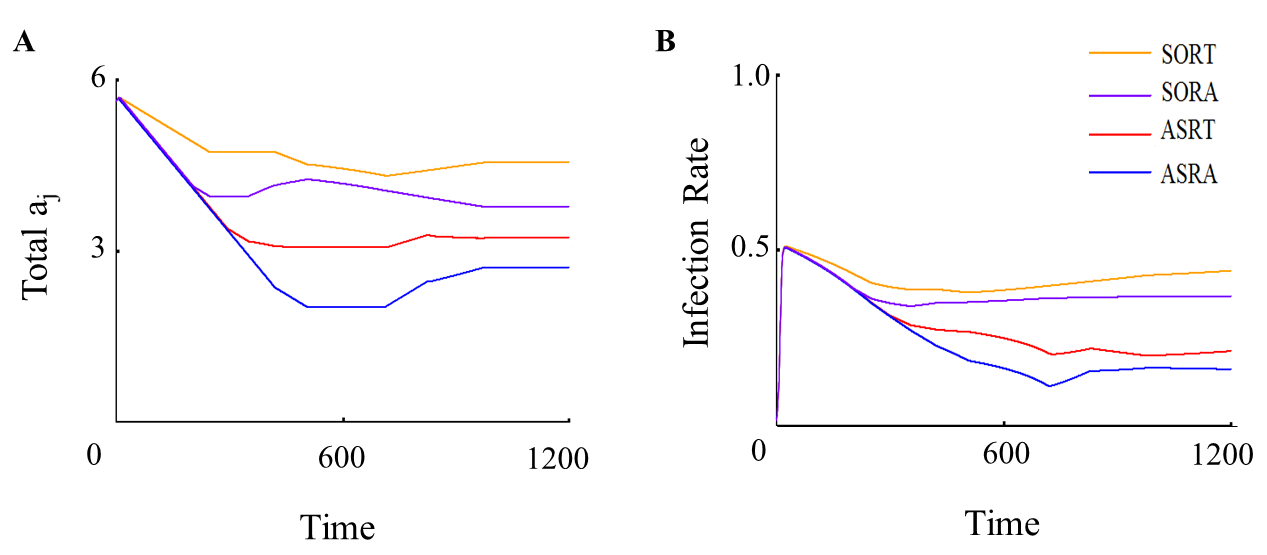}
%\hspace{1mm}
\caption{Illustration of the dynamics of the system over time. A. The socialization rates of all groups decline initially before stabilising at different levels, related to B. corresponding changes in the prevalence of infection in different groups. Model parameters are set as $\sigma_l = 4, \sigma_h = 12, \rho_l = 30, \rho_h = 60, r = 1$. Line colors designate the different groups as indicated in the figure. ASRT is asocial risk-tolerant, ASRA is asocial risk-averse, SORT is social risk-tolerant, SORA is social-risk averse.}  
\label{BasicSim}
\end{figure}

We simulate the model under a variety of parameters and find a number of consistent patterns.  Figure \ref{BasicSim} shows the total socialization of players within each group over time for one instance of the simulation.

%\begin{figure}[h]
%\includegraphics[scale = 0.25]{Figures/history_graph_4stack_v3.png}
%\hspace{1mm}
%\includegraphics[scale = 0.25]{Figures/history_graph_inf_v3.png}
%\caption{Illustration of the dynamics of the system over time. a) The socialization rates of all groups decline initially before stabilising at different levels, related to b) corresponding changes in the prevalence of infection in different groups. Model parameters are set as $\sigma_l = 4, \sigma_h = 12, \rho_l = 30, \rho_h = 60, r = 1$. Line colors designate the different groups as indicated in the figure.}  
%\label{BasicSim}
%\end{figure}

\subsection*{Neutral Re-assortment of Social Contacts}
In its most extreme expression, individuals were allowed to abandon all existing social contact with individuals who did not maximize their individual joint utility in both socializing and risk tolerance. Under this scenario, $r=1$ and we see that players settle into an equilibrium level of socializing and infection, and that both levels are higher for groups with higher $\sigma_j$ and lower $\rho_j$.  The utility gained from socializing is subject to two levels of negative feedback.  In the short term, higher levels of socialization receive less marginal utility due to the logarithm, meaning that increasing socialization is of less benefit to the more social groups.  In the long term, higher levels of socialization result in higher levels of infection in the population, increasing the costs of socializing.  If we take the derivative of $u_j$ with respect to $a'_{j,k}$ and rearrange, we find that

$$
\frac{\partial u}{\partial a'_{j,k}} > 0 \mbox{ iff }   a'_{j,k} < \frac{\sigma_j}{\rho_j \beta I_k}.
$$

Thus we see more socialization in groups with higher $\sigma_j$ and lower $\rho_j$.

Simulations of this model repeatedly show the emergence of perfect homophily as time progresses.  Each group segregates from the others and individuals only socialize within their own group.

\begin{figure}[h]
\includegraphics[scale = 0.35]{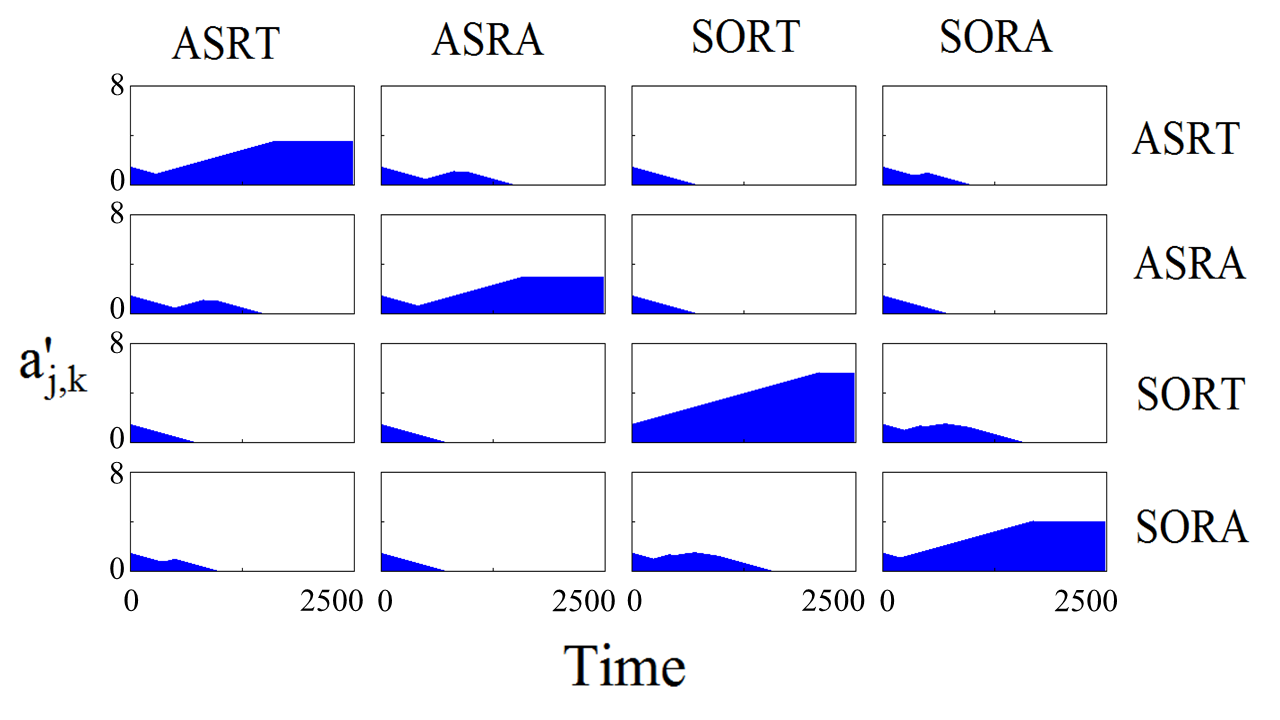}
\caption{Within and between group socialization rates over time with $r = 1$. Within group socialization rates increase over time until the system stabilises with perfect homophily according to group membership. Model parameters are set as $\sigma_l = 4, \sigma_h = 12, \rho_l = 30, \rho_h = 60$. ASRT is asocial risk-tolerant, ASRA is asocial risk-averse, SORT is social risk-tolerant, SORA is social risk-averse.}
\label{PerfectHomophily}
\end{figure}

Figure \ref{PerfectHomophily} shows how these social patterns emerge over time.  The initial epidemic drives all socialization levels down, which in turn causes the infection to recede.  However, when socialization levels recover, gains in socialization primarily occur within each group due to a natural stratification of epidemiological risk.

The differing levels of infection within each group mean that only more social players are willing to tolerate higher levels of infection to socialize more, so they begin increasing $a$ sooner after the initial epidemic, and stably socialize more at equilibrium.  As a result, other players are less willing to socialize with those players in particular. Because the positive term in each player's utility function does not vary based on player subgroup, socializing with less-infected subgroups yields the same payoff for a smaller risk.

Therefore, everyone wants to socialize with the safest, least social group - the one with high $\rho_j$ and low $\sigma_j$ (ASRA).  However, everyone in this group also prefers to socialize with players from their own group.  Consequently, players from least social and most risk averse group can increase their utility by decreasing all $a_{2,k}$ to $0$, where $k \neq 2$, and increasing $a_{2,2}$ to compensate.  This lowers their risk while still allowing them to benefit from socializing.  And because $a'_{2,2} = a_{2,2}$, they are able to socialize at exactly their preferred level, rendering other socialization with other groups unnecessary.

The continuation of the above mechanism propagated through successively more social and less risk averse groups recursively creates a linear hierarchy in the population.  Every other group increases $a_{j,2}$, but this is not reciprocated, so $a'_{j,2} = 0$.  The least infected group has effectively removed itself from the population  Without the ability to socialize with their preferred group, they have to increase other $a_{j,k}$.  The second least infected group (group $1$ in the figure) then becomes the preferred social partner of each group, including themselves.  As the same process repeats recursively, each group in turn segregates itself from the more infected groups and achieves its equilibrium socialization level internally.  Eventually, all groups end up socializing internally, with more infected groups unilaterally being rebuffed in their attempts to socialize more by increasing their socialization with less infected groups ($a_{j,k} > 0$, but $a_{k,j} = 0$).

\subsection*{Retention of Preexisting Social Contacts}

The perfect homophily generated above results from a scenario in which individuals drop all of their existing social relationships from outside of their group and replace them with individuals from their own group.  This assumption is unrealistic because individuals value aspects of their relationships other than the risk of infection, or may also be constrained in their choice of associations. This means that relationships are not perfectly replaceable, with each potential partner equivalently able to satisfy the desire for social contact.  We therefore also explore an adjusted scenario for our model to account for some relationships being more valuable by increasing the benefits of heterogeneous socialization. In this scenario, we assume that possible social relationships and activities have varying values, and that players will drop the least valuable ones first when possible within a group, meaning the benefits obtained by socializing are limited if a player adheres too closely within any one group.

\begin{figure}[h]
\includegraphics[scale = 0.35]{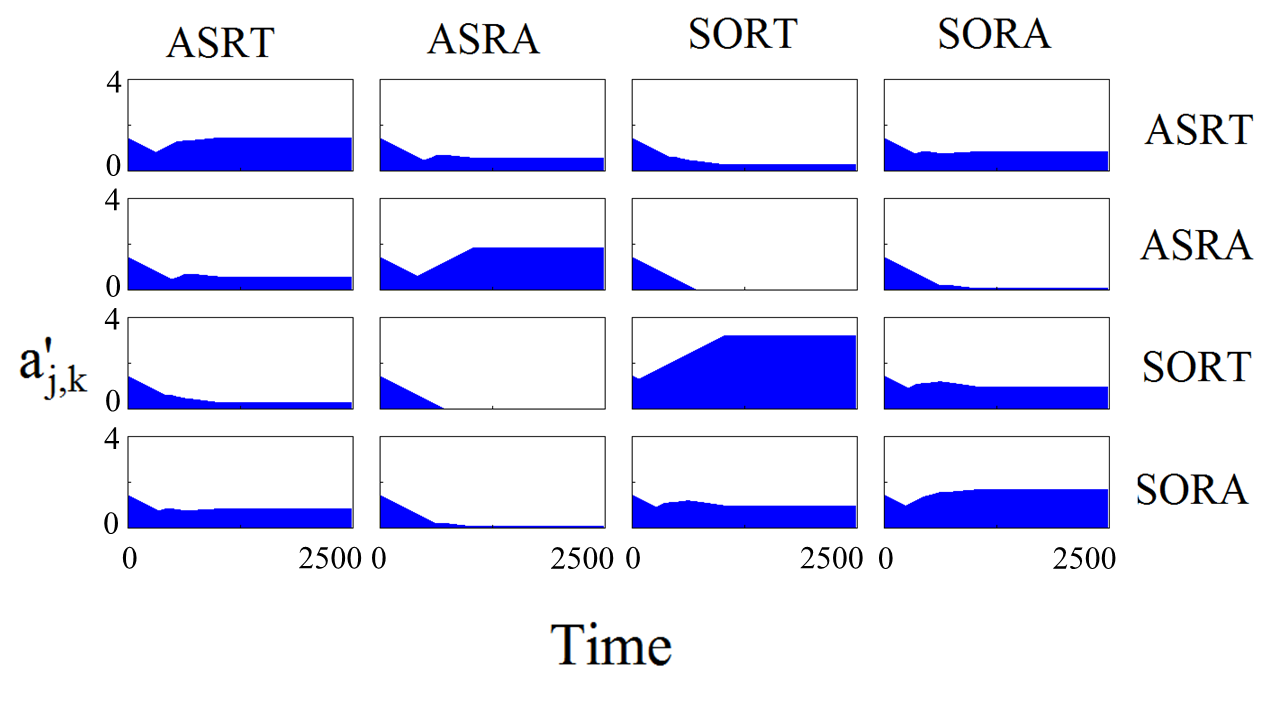}
\caption{Within and between group socialization rates over time with $r = 1/2$ and $m = 3$. Within group socialization rates increase over time until the system stabilises with partial homophily according to group membership. Other model parameters are set as $\sigma_l = 4, \sigma_h = 12, \rho_l = 30, \rho_h = 60$. ASRT is asocial risk-tolerant, ASRA is asocial risk-averse, SORT is social risk-tolerant, SORA is social risk-averse.}
\label{PartialHomophily}
\end{figure}

Figure \ref{PartialHomophily} shows the pairwise socialization between members of each group over time that result from adding these social constraints.  We see now see only partial homophily emerging. Players still socialize primarily within their own group at equilibrium, but still maintain certain associations with other groups as the benefits from the multiplier outweigh the increased infection risks.  Pairs with more similar home-group infection risks still tend to socialize more than dissimilar pairs.

\begin{figure}[h]
\includegraphics[scale = 0.3]{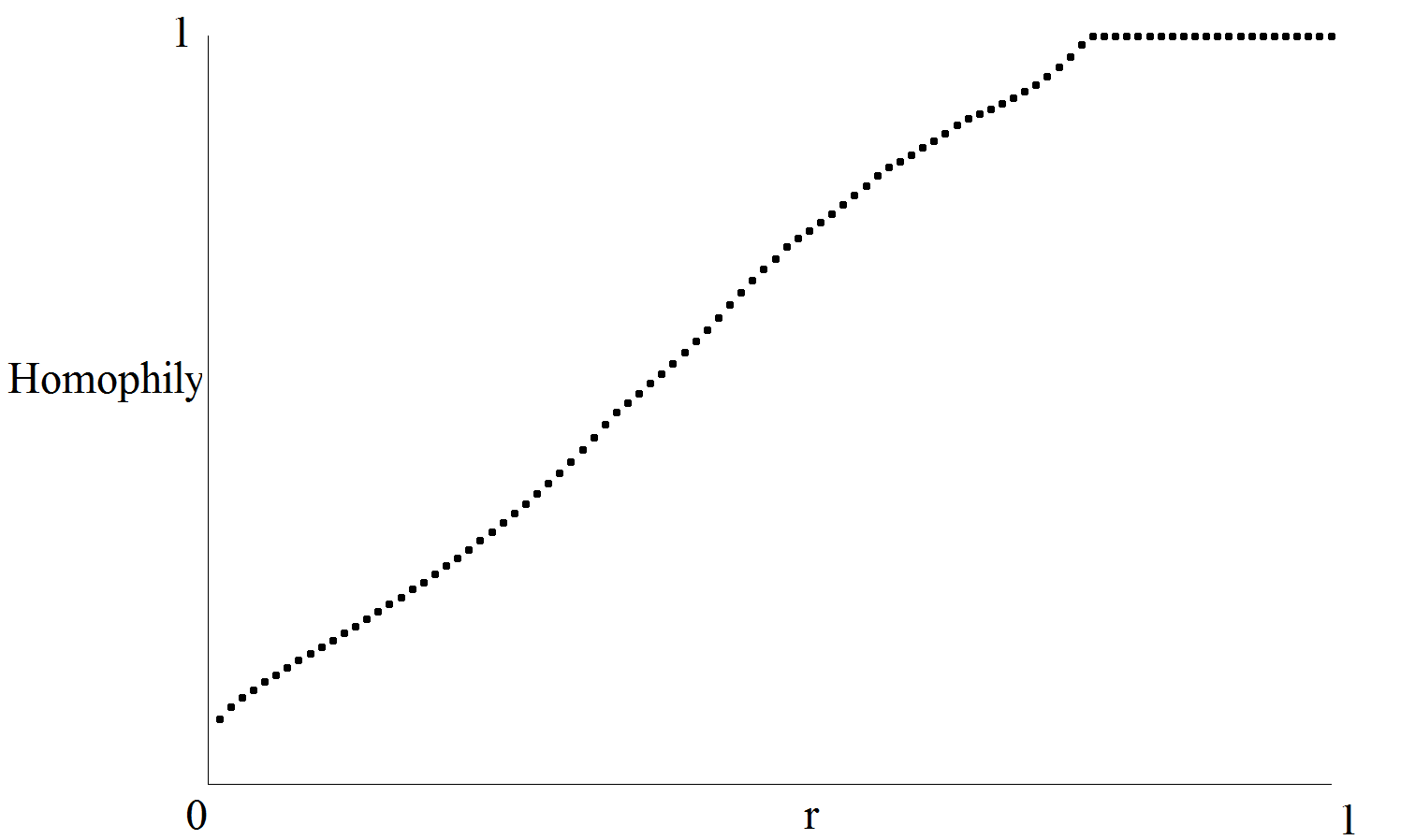}
\caption{Homophily in socialization rates according to group membership as a function of $r$. Other model parameters are set as $\sigma_l = 4, \sigma_h = 12, \rho_l = 30, \rho_h = 60$.}
\label{PartialHomophily2}
\end{figure}

We can vary the value individuals place on particular, irreplaceable contacts by adjusting the steepness of the multiplier function (Fig. \ref{PartialHomophily2}).   At low values of $r$ (less steep multiplier function) we see low levels of homophily as in Fig. \ref{PerfectHomophily}. However, as we increase $r$ we see that mixing between groups decreases until perfect homophily emerges at values of $r$ close to one.

\subsection*{Diversity as a Driver of Homophily}
Here we demonstrate that the behavioural response of a population to an infectious outbreak can drive the emergence of homophily as a result of individual decision-making strategies. Explaining emergent organizational structures that impact the function and efficiency of populations, but arise solely from individual behaviors, is a fundamental challenge in understanding social systems \cite{kulig2015modelling, newman2001best,hill2011co,smith2003complex, hock2010systems}. The role of infectious disease is of particular interest due to both the successful function of a society (a benefit) and successful transmission of infection (a cost) relying on similar types of contacts and behaviors \cite{hock2012social,udiani2020disease, romano2021tradeoff, youssef2013mitigation}. Changing our assumptions to assign value to pre-existing relationships lessens, but does not remove, the emergence of homophily. 

In our model, the unanticipated emergence of homophilous groups along axes prioritizing risk aversion or social interactions parallels experimental and theoretical work emphasizing risk aversion as a key component of the structural emergence of homophily \cite{currarini2016identity, kovavrik2014risk}. Uncertainty of the environment and individual differences in risk aversion have been posited to interact for the generation of network structure \cite{kovavrik2014risk}. Interestingly, our results are similar, but achieve these similarities without making the previously included assumption of existing social structure. Even in our extended model that places value on existing relationships, homophilous ties emerge in response to a system-wide outbreak when both risk and risk perception are driven by behavior. This shows that large epidemics have the capacity to disturb and reorganize social structure along axes of risk and social preference, as evidenced during the HIV and COVID-19 pandemics \cite{buck2020ecological,leigh1989reasons,ellen2002improving,dupas2011teenagers}. Such population level emergence of assortative mixing has important consequences for anticipating the spread of disease in a heterogeneous society, especially as homophily will also increase modularity and alter the dynamics of the spread of both disease and information \cite{fefferman2021homophily, jackson2013diffusion, nunn2015infectious,sah2017unraveling}.

Perhaps most intriguingly, however, our findings suggest that observable emergent community structure may arise from individual differences in categorical preference or assessment of the risks and benefits of social contact over time. While disease is an important example of socially contagious risk, it is certainly not the only one. Cultural norms and the perceived threat of homogenization eroding group identity \cite{falomir2004perceived,zarate2004cultural} may similarly act as a driving factor in constructing and maintaining social divisions among groups. Of course, individual game theoretic perspectives are not the only proposed mechanism for the emergence of such structures \cite{jackson2019behavioral,elder2007emergence,gulati2012rise, newman2002random}. However, our results contribute to understanding how simple, individual perceptions and behaviors may yield highly organized, and operationally beneficial, global outcomes. 

\subsection*{Implications for Public Health}

Emergent structure in society that arises out of individual needs to balance conflicting goals, can provide critical insight into how to influence the individual perceptions and behaviors from which they are formed. In the case of infectious disease outbreaks in human populations, the natural self-organization into homophilous groups offers immediate potential routes for public health intervention. Understanding the independent value propositions that drive community formation can allow the design of strategies that, while marginally less effective in the absolute reduction of infection risk, achieve meaningful reduction without incurring the same social costs. By considering self-organization rooted in multi-factorial utility we can begin to produce useful, quantitative tools to inform policy and improve real-world adoption of mitigation strategies. 

\subsection*{Acknowledgements}
We gratefully acknowledge helpful discussions with collaborator R.A. Bentley and S. Carrignon, and funding support for this work from NSF DEB 2028710.

\subsection*{Code Availability}
Project code is available on the code hosting platform GitHub at\\ https://github.com/kazarraha/SocDistModel

\bibliography{References}

\begin{thebibliography}{63}
\providecommand{\natexlab}[1]{#1}
\providecommand{\url}[1]{\texttt{#1}}
\expandafter\ifx\csname urlstyle\endcsname\relax
  \providecommand{\doi}[1]{doi: #1}\else
  \providecommand{\doi}{doi: \begingroup \urlstyle{rm}\Url}\fi

\bibitem[Fu et~al.(2012)Fu, Nowak, Christakis, and Fowler]{fu2012evolution}
Feng Fu, Martin~A Nowak, Nicholas~A Christakis, and James~H Fowler.
\newblock The evolution of homophily.
\newblock \emph{Scientific reports}, 2\penalty0 (1):\penalty0 1--6, 2012.

\bibitem[Currarini et~al.(2016)Currarini, Matheson, and
  Vega-Redondo]{currarini2016simple}
Sergio Currarini, Jesse Matheson, and Fernando Vega-Redondo.
\newblock A simple model of homophily in social networks.
\newblock \emph{European Economic Review}, 90:\penalty0 18--39, 2016.

\bibitem[Treur(2019)]{treur2019mathematical}
Jan Treur.
\newblock Mathematical analysis of the emergence of communities based on
  coevolution of social contagion and bonding by homophily.
\newblock \emph{Applied Network Science}, 4\penalty0 (1):\penalty0 1--30, 2019.

\bibitem[Holme and Newman(2006)]{holme2006coevolution}
Petter Holme and M.~E.~J. Newman.
\newblock Nonequilibrium phase transition in the coevolution of networks and
  opinions.
\newblock \emph{Phys. Rev. E}, 74:\penalty0 056108, Nov 2006.
\newblock \doi{10.1103/PhysRevE.74.056108}.

\bibitem[Pearson et~al.(2006)Pearson, Steglich, and
  Snijders]{pearson2006homophily}
Michael Pearson, Christian Steglich, and Tom Snijders.
\newblock Homophily and assimilation among sport-active adolescent substance
  users.
\newblock \emph{Connections}, 27\penalty0 (1):\penalty0 47--63, 2006.

\bibitem[McPherson et~al.(2001)McPherson, Smith-Lovin, and
  Cook]{mcpherson2001homophily}
Miller McPherson, Lynn Smith-Lovin, and James~M Cook.
\newblock Birds of a feather: homophily in social networks.
\newblock \emph{Annual Review of Sociology}, 27:\penalty0 415--444, 2001.

\bibitem[Sah et~al.(2017)Sah, Leu, Cross, Hudson, and
  Bansal]{sah2017unraveling}
Pratha Sah, Stephan~T Leu, Paul~C Cross, Peter~J Hudson, and Shweta Bansal.
\newblock Unraveling the disease consequences and mechanisms of modular
  structure in animal social networks.
\newblock \emph{Proceedings of the National Academy of Sciences}, 114\penalty0
  (16):\penalty0 4165--4170, 2017.

\bibitem[Salath{\'e} and Jones(2010)]{salathe2010dynamics}
Marcel Salath{\'e} and James~H Jones.
\newblock Dynamics and control of diseases in networks with community
  structure.
\newblock \emph{PLoS computational biology}, 6\penalty0 (4):\penalty0 e1000736,
  2010.

\bibitem[Evans et~al.(2021)Evans, Hodgson, Boogert, and Silk]{evans2021group}
Julian~C Evans, David~J Hodgson, Neeltje~J Boogert, and Matthew~J Silk.
\newblock Group size and modularity interact to shape the spread of infection
  and information through animal societies.
\newblock \emph{bioRxiv}, 2021.

\bibitem[Cinelli et~al.(2021)Cinelli, Morales, Galeazzi, Quattrociocchi, and
  Starnini]{cinelli2021echo}
Matteo Cinelli, Gianmarco De~Francisci Morales, Alessandro Galeazzi, Walter
  Quattrociocchi, and Michele Starnini.
\newblock The echo chamber effect on social media.
\newblock \emph{Proceedings of the National Academy of Sciences}, 118\penalty0
  (9), 2021.

\bibitem[Leppin and Aro(2009)]{leppin2009risk}
Anja Leppin and Arja~R Aro.
\newblock Risk perceptions related to sars and avian influenza: theoretical
  foundations of current empirical research.
\newblock \emph{International journal of behavioral medicine}, 16\penalty0
  (1):\penalty0 7--29, 2009.

\bibitem[Schaller(2011)]{schaller2011behavioural}
Mark Schaller.
\newblock The behavioural immune system and the psychology of human sociality.
\newblock \emph{Philosophical Transactions of the Royal Society B: Biological
  Sciences}, 366\penalty0 (1583):\penalty0 3418--3426, 2011.

\bibitem[Schaller(2006)]{schaller2006parasites}
Mark Schaller.
\newblock Parasites, behavioral defenses, and the social psychological
  mechanisms through which cultures are evoked.
\newblock \emph{Psychological Inquiry}, 17\penalty0 (2):\penalty0 96--101,
  2006.

\bibitem[Collier et~al.(2020)Collier, Cotten, and Roush]{collier2020risk}
Trevor Collier, Stephen~J Cotten, and Justin Roush.
\newblock Risk aversion, guilt, and pandemic behavior.
\newblock \emph{Guilt, and Pandemic Behavior (December 1, 2020)}, 2020.

\bibitem[Baumeister and Leary(1995)]{baumeister1995need}
Roy~F Baumeister and Mark~R Leary.
\newblock The need to belong: desire for interpersonal attachments as a
  fundamental human motivation.
\newblock \emph{Psychological bulletin}, 117\penalty0 (3):\penalty0 497, 1995.

\bibitem[Hawkley and Cacioppo(2010)]{hawkley2010loneliness}
Louise~C Hawkley and John~T Cacioppo.
\newblock Loneliness matters: A theoretical and empirical review of
  consequences and mechanisms.
\newblock \emph{Annals of behavioral medicine}, 40\penalty0 (2):\penalty0
  218--227, 2010.

\bibitem[Eisenberger and Lieberman(2004)]{eisenberger2004rejection}
Naomi~I Eisenberger and Matthew~D Lieberman.
\newblock Why rejection hurts: a common neural alarm system for physical and
  social pain.
\newblock \emph{Trends in cognitive sciences}, 8\penalty0 (7):\penalty0
  294--300, 2004.

\bibitem[Sawada et~al.(2018)Sawada, Auger, and Lydon]{sawada2018activation}
Natsumi Sawada, Emilie Auger, and John~E Lydon.
\newblock Activation of the behavioral immune system: Putting the brakes on
  affiliation.
\newblock \emph{Personality and Social Psychology Bulletin}, 44\penalty0
  (2):\penalty0 224--237, 2018.

\bibitem[Sacco et~al.(2014)Sacco, Young, and Hugenberg]{sacco2014balancing}
Donald~F Sacco, Steven~G Young, and Kurt Hugenberg.
\newblock Balancing competing motives: Adaptive trade-offs are necessary to
  satisfy disease avoidance and interpersonal affiliation goals.
\newblock \emph{Personality and Social Psychology Bulletin}, 40\penalty0
  (12):\penalty0 1611--1623, 2014.

\bibitem[Biswas et~al.(2020)Biswas, Rahaman, Biswas, Haque, and
  Ibrahim]{biswas2020effects}
Mohitosh Biswas, Shawonur Rahaman, Tapash~Kumar Biswas, Zahirul Haque, and
  Baharudin Ibrahim.
\newblock Effects of sex, age and comorbidities on the risk of infection and
  death associated with covid-19: a meta-analysis of 47807 confirmed cases.
\newblock \emph{Age and Comorbidities on the Risk of Infection and Death
  Associated with COVID-19: A Meta-Analysis of}, 47807, 2020.

\bibitem[Maggini et~al.(2018)Maggini, Pierre, and Calder]{maggini2018immune}
Silvia Maggini, Adeline Pierre, and Philip~C Calder.
\newblock Immune function and micronutrient requirements change over the life
  course.
\newblock \emph{Nutrients}, 10\penalty0 (10):\penalty0 1531, 2018.

\bibitem[Caligiuri and Tarique(2012)]{caligiuri2012dynamic}
Paula Caligiuri and Ibraiz Tarique.
\newblock Dynamic cross-cultural competencies and global leadership
  effectiveness.
\newblock \emph{Journal of world Business}, 47\penalty0 (4):\penalty0 612--622,
  2012.

\bibitem[Lauriola et~al.(2007)Lauriola, Levin, and Hart]{lauriola2007common}
Marco Lauriola, Irwin~P Levin, and Stephanie~S Hart.
\newblock Common and distinct factors in decision making under ambiguity and
  risk: A psychometric study of individual differences.
\newblock \emph{Organizational Behavior and Human Decision Processes},
  104\penalty0 (2):\penalty0 130--149, 2007.

\bibitem[Petrocchi et~al.(2021)Petrocchi, Iannello, Ongaro, Antonietti, and
  Pravettoni]{petrocchi2021interplay}
S~Petrocchi, P~Iannello, G~Ongaro, A~Antonietti, and G~Pravettoni.
\newblock The interplay between risk and protective factors during the initial
  height of the covid-19 crisis in italy: The role of risk aversion and
  intolerance of ambiguity on distress.
\newblock \emph{Current Psychology}, pages 1--12, 2021.

\bibitem[Blanchard et~al.(2011)Blanchard, Griebel, Pobbe, and
  Blanchard]{blanchard2011riskassess}
D.~Caroline Blanchard, Guy Griebel, Roger Pobbe, and Robert~J Blanchard.
\newblock Risk assessment as an evolved threat detection and analysis process.
\newblock \emph{Neuroscience \& Biobehavioral Reviews}, 35\penalty0
  (4):\penalty0 991--998, 2011.

\bibitem[Smith and Blumstein(2008)]{smith2008fitness}
Brian~R Smith and Daniel~T Blumstein.
\newblock Fitness consequences of personality: a meta-analysis.
\newblock \emph{Behavioral Ecology}, 19\penalty0 (2):\penalty0 448--455, 2008.

\bibitem[van Oers et~al.(2004)van Oers, Drent, Goede, and van
  Noordwijk]{vanoers2003heritab}
Kees van Oers, Piet~J Drent, Piet~de Goede, and Arie~J van Noordwijk.
\newblock Realized heritability and repeatability of risk-taking behaviour in
  relation to avian personalities.
\newblock \emph{Proceedings of the Royal Society of London B: Biological
  Sciences}, 271\penalty0 (1534):\penalty0 65--73, 2004.

\bibitem[Cesarini et~al.(2009)Cesarini, Dawes, Johannesson, Lichtenstein, and
  Wallace]{cesarini2009genetic}
David Cesarini, Christopher~T Dawes, Magnus Johannesson, Paul Lichtenstein, and
  Bj{\"o}rn Wallace.
\newblock Genetic variation in preferences for giving and risk taking.
\newblock \emph{The Quarterly Journal of Economics}, 124\penalty0 (2):\penalty0
  809--842, 2009.

\bibitem[Zuckerman and Kuhlman(2000)]{zuckerman2001risk-taking}
M~Zuckerman and D~M Kuhlman.
\newblock Personality and risk-taking: Common biosocial factors.
\newblock \emph{Journal of Personality}, 68\penalty0 (6):\penalty0 999--1029,
  2000.

\bibitem[Nicholson et~al.(2005)Nicholson, Soane, Fenton‐O'Creevy, and
  Willman]{nicholson2005persandrisk}
Nigel Nicholson, Emma Soane, Mark Fenton‐O'Creevy, and Paul Willman.
\newblock Personality and domain‐specific risk taking.
\newblock \emph{Journal of Risk Research}, 8\penalty0 (2):\penalty0 157--176,
  2005.

\bibitem[Van~Bavel et~al.(2020)Van~Bavel, Baicker, Boggio, Capraro, Cichocka,
  Cikara, Crockett, Crum, Douglas, Druckman, et~al.]{van2020using}
Jay~J Van~Bavel, Katherine Baicker, Paulo~S Boggio, Valerio Capraro, Aleksandra
  Cichocka, Mina Cikara, Molly~J Crockett, Alia~J Crum, Karen~M Douglas,
  James~N Druckman, et~al.
\newblock Using social and behavioural science to support covid-19 pandemic
  response.
\newblock \emph{Nature human behaviour}, 4\penalty0 (5):\penalty0 460--471,
  2020.

\bibitem[Paget(2009)]{pagetinfluenza}
John Paget.
\newblock The influenza pandemic and europe: the social impact and public
  health.
\newblock \emph{Italian Journal of Public Health}, 6:\penalty0 257, 2009.

\bibitem[Fenichel(2013)]{fenichel2013economic}
Eli~P Fenichel.
\newblock Economic considerations for social distancing and behavioral based
  policies during an epidemic.
\newblock \emph{Journal of health economics}, 32\penalty0 (2):\penalty0
  440--451, 2013.

\bibitem[Bell and Gersbach(2009)]{bell2009macroeconomics}
Clive Bell and Hans Gersbach.
\newblock The macroeconomics of targeting: the case of an enduring epidemic.
\newblock \emph{Journal of Health Economics}, 28\penalty0 (1):\penalty0 54--72,
  2009.

\bibitem[Currarini and Mengel(2016)]{currarini2016identity}
Sergio Currarini and Friederike Mengel.
\newblock Identity, homophily and in-group bias.
\newblock \emph{European Economic Review}, 90:\penalty0 40--55, 2016.

\bibitem[Kov{\'a}{\v{r}}{\'\i}k and Van~der Leij(2014)]{kovavrik2014risk}
Jarom{\'\i}r Kov{\'a}{\v{r}}{\'\i}k and Marco~J Van~der Leij.
\newblock Risk aversion and social networks.
\newblock \emph{Review of Network Economics}, 13\penalty0 (2):\penalty0
  121--155, 2014.

\bibitem[Arregui et~al.(2018)Arregui, Aleta, Sanz, and
  Moreno]{arregui2018projecting}
Sergio Arregui, Alberto Aleta, Joaqu{\'\i}n Sanz, and Yamir Moreno.
\newblock Projecting social contact matrices to different demographic
  structures.
\newblock \emph{PLoS computational biology}, 14\penalty0 (12):\penalty0
  e1006638, 2018.

\bibitem[Anderson and May(1992)]{anderson1992infectious}
Roy~M Anderson and Robert~M May.
\newblock \emph{Infectious diseases of humans: dynamics and control}.
\newblock Oxford university press, 1992.

\bibitem[Catching et~al.(2021)Catching, Capponi, Yeh, Bianco, and
  Andino]{catching2021examining}
Adam Catching, Sara Capponi, Ming~Te Yeh, Simone Bianco, and Raul Andino.
\newblock Examining the interplay between face mask usage, asymptomatic
  transmission, and social distancing on the spread of covid-19.
\newblock \emph{Scientific reports}, 11\penalty0 (1):\penalty0 1--11, 2021.

\bibitem[Cetrulo et~al.(2020)Cetrulo, Guarascio, and
  Virgillito]{cetrulo2020privilege}
Armanda Cetrulo, Dario Guarascio, and Maria~Enrica Virgillito.
\newblock The privilege of working from home at the time of social distancing.
\newblock \emph{Intereconomics}, 55:\penalty0 142--147, 2020.

\bibitem[Townsend et~al.(2020)Townsend, Hawley, Stephenson, and
  Williams]{townsend2020emerging}
Andrea~K Townsend, Dana~M Hawley, Jessica~F Stephenson, and Keelah~EG Williams.
\newblock Emerging infectious disease and the challenges of social distancing
  in human and non-human animals.
\newblock \emph{Proceedings of the Royal Society B}, 287\penalty0
  (1932):\penalty0 20201039, 2020.

\bibitem[Kulig et~al.(2015)Kulig, Drozdz, Kwapien, and
  Oswiecimka]{kulig2015modelling}
Andrzej Kulig, Stanislaw Drozdz, Jaroslaw Kwapien, and Pawel Oswiecimka.
\newblock Modelling subtle growth of linguistic networks.
\newblock \emph{Phys. Rev. E}, 91:\penalty0 032810, 2015.

\bibitem[Newman(2001)]{newman2001best}
Mark~EJ Newman.
\newblock Who is the best connected scientist? a study of scientific
  coauthorship networks.
\newblock \emph{Phys. Rev. E}, 64\penalty0 (016131), 2001.

\bibitem[Hill et~al.(2011)Hill, Walker, Bo{\v{z}}i{\v{c}}evi{\'c}, Eder,
  Headland, Hewlett, Hurtado, Marlowe, Wiessner, and Wood]{hill2011co}
Kim~R Hill, Robert~S Walker, Miran Bo{\v{z}}i{\v{c}}evi{\'c}, James Eder,
  Thomas Headland, Barry Hewlett, A~Magdalena Hurtado, Frank Marlowe, Polly
  Wiessner, and Brian Wood.
\newblock Co-residence patterns in hunter-gatherer societies show unique human
  social structure.
\newblock \emph{science}, 331\penalty0 (6022):\penalty0 1286--1289, 2011.

\bibitem[Smith et~al.(2003)Smith, Brighton, and Kirby]{smith2003complex}
Kenny Smith, Henry Brighton, and Simon Kirby.
\newblock Complex systems in language evolution: the cultural emergence of
  compositional structure.
\newblock \emph{Advances in complex systems}, 6\penalty0 (04):\penalty0
  537--558, 2003.

\bibitem[Hock et~al.(2010)Hock, Ng, and Fefferman]{hock2010systems}
Karlo Hock, Kah~Loon Ng, and Nina~H Fefferman.
\newblock Systems approach to studying animal sociality: Individual position
  versus group organization in dynamic social network models.
\newblock \emph{PLoS One}, 5\penalty0 (12):\penalty0 e15789, 2010.

\bibitem[Hock and Fefferman(2012)]{hock2012social}
Karlo Hock and Nina~H Fefferman.
\newblock Social organization patterns can lower disease risk without
  associated disease avoidance or immunity.
\newblock \emph{Ecological Complexity}, 12:\penalty0 34--42, 2012.

\bibitem[Udiani and Fefferman(2020)]{udiani2020disease}
Oyita Udiani and Nina~H Fefferman.
\newblock How disease constrains the evolution of social systems.
\newblock \emph{Proceedings of the Royal Society B}, 287\penalty0
  (1932):\penalty0 20201284, 2020.

\bibitem[Romano et~al.(2021)Romano, Sueur, and MacIntosh]{romano2021tradeoff}
Val{\'e}ria Romano, C{\'e}dric Sueur, and Andrew~JJ MacIntosh.
\newblock The tradeoff between information and pathogen transmission in animal
  societies.
\newblock \emph{Oikos}, 2021.

\bibitem[Youssef and Scoglio(2013)]{youssef2013mitigation}
Mina Youssef and Caterina Scoglio.
\newblock Mitigation of epidemics in contact networks through optimal contact
  adaptation.
\newblock \emph{Mathematical biosciences and engineering: MBE}, 10\penalty0
  (4):\penalty0 1227, 2013.

\bibitem[Buck and Weinstein(2020)]{buck2020ecological}
Julia~C Buck and Sara~B Weinstein.
\newblock The ecological consequences of a pandemic.
\newblock \emph{Biology Letters}, 16\penalty0 (11):\penalty0 20200641, 2020.

\bibitem[Leigh(1989)]{leigh1989reasons}
Barbara~Critchlow Leigh.
\newblock Reasons for having and avoiding sex: Gender, sexual orientation, and
  relationship to sexual behavior.
\newblock \emph{Journal of Sex Research}, 26\penalty0 (2):\penalty0 199--209,
  1989.

\bibitem[Ellen et~al.(2002)Ellen, Adler, Gurvey, Dunlop, Millstein, and
  Tschann]{ellen2002improving}
Jonathan~M Ellen, Nancy Adler, Jill~E Gurvey, Miranda~BV Dunlop, Susan~G
  Millstein, and Jeanne Tschann.
\newblock Improving predictions of condom behavioral intentions with
  partner-specific measures of risk perception.
\newblock \emph{Journal of Applied Social Psychology}, 32\penalty0
  (3):\penalty0 648--663, 2002.

\bibitem[Dupas(2011)]{dupas2011teenagers}
Pascaline Dupas.
\newblock Do teenagers respond to hiv risk information? evidence from a field
  experiment in kenya.
\newblock \emph{American Economic Journal: Applied Economics}, 3\penalty0
  (1):\penalty0 1--34, 2011.

\bibitem[Fefferman et~al.(2021)Fefferman, Silk, Pasquale, and
  Moody]{fefferman2021homophily}
Nina~H Fefferman, Matthew~J Silk, Dana~K Pasquale, and James Moody.
\newblock Homophily in risk and behavior complicate understanding the covid-19
  epidemic curve.
\newblock \emph{medRxiv}, 2021.

\bibitem[Jackson and L{\'o}pez-Pintado(2013)]{jackson2013diffusion}
Matthew~O Jackson and Dunia L{\'o}pez-Pintado.
\newblock Diffusion and contagion in networks with heterogeneous agents and
  homophily.
\newblock \emph{Network Science}, 1\penalty0 (1):\penalty0 49--67, 2013.

\bibitem[Nunn et~al.(2015)Nunn, Jord{\'a}n, McCabe, Verdolin, and
  Fewell]{nunn2015infectious}
Charles~L Nunn, Ferenc Jord{\'a}n, Collin~M McCabe, Jennifer~L Verdolin, and
  Jennifer~H Fewell.
\newblock Infectious disease and group size: more than just a numbers game.
\newblock \emph{Philosophical Transactions of the Royal Society B: Biological
  Sciences}, 370\penalty0 (1669):\penalty0 20140111, 2015.

\bibitem[Falomir-Pichastor et~al.(2004)Falomir-Pichastor, Mu{\~n}oz-Rojas,
  Invernizzi, and Mugny]{falomir2004perceived}
Juan~Manuel Falomir-Pichastor, Daniel Mu{\~n}oz-Rojas, Federica Invernizzi, and
  Gabriel Mugny.
\newblock Perceived in-group threat as a factor moderating the influence of
  in-group norms on discrimination against foreigners.
\newblock \emph{European Journal of Social Psychology}, 34\penalty0
  (2):\penalty0 135--153, 2004.

\bibitem[Zarate et~al.(2004)Zarate, Garcia, Garza, and
  Hitlan]{zarate2004cultural}
Michael~A Zarate, Berenice Garcia, Azenett~A Garza, and Robert~T Hitlan.
\newblock Cultural threat and perceived realistic group conflict as dual
  predictors of prejudice.
\newblock \emph{Journal of experimental social psychology}, 40\penalty0
  (1):\penalty0 99--105, 2004.

\bibitem[Jackson and Storms(2019)]{jackson2019behavioral}
Matthew~O Jackson and Evan Storms.
\newblock Behavioral communities and the atomic structure of networks.
\newblock \emph{Available at SSRN 3049748}, 2019.

\bibitem[ELDER-VASS(2007)]{elder2007emergence}
DAVE ELDER-VASS.
\newblock For emergence: refining archer's account of social structure.
\newblock \emph{Journal for the Theory of Social Behaviour}, 37\penalty0
  (1):\penalty0 25--44, 2007.

\bibitem[Gulati et~al.(2012)Gulati, Sytch, and Tatarynowicz]{gulati2012rise}
Ranjay Gulati, Maxim Sytch, and Adam Tatarynowicz.
\newblock The rise and fall of small worlds: Exploring the dynamics of social
  structure.
\newblock \emph{Organization Science}, 23\penalty0 (2):\penalty0 449--471,
  2012.

\bibitem[Newman et~al.(2002)Newman, Watts, and Strogatz]{newman2002random}
Mark~EJ Newman, Duncan~J Watts, and Steven~H Strogatz.
\newblock Random graph models of social networks.
\newblock \emph{Proceedings of the national academy of sciences}, 99\penalty0
  (suppl 1):\penalty0 2566--2572, 2002.

\end{thebibliography}
\end{document}